# Pediatric hospitalizations associated with respiratory syncytial virus (RSV) and influenza, and the frequency of asthma as a secondary diagnosis

Edward Goldstein [1]

1) London, UK. edmugo3@gmail.com

## Abstract

Background: There is uncertainty about the burden of severe outcomes associated with RSV and influenza in different subgroups of children defined by age, as well as age and presence of underlying health conditions.

Methods: We have applied previously developed methodology to estimate the rates of influenza and RSV-associated hospitalizations with different diagnoses in different age subgroups of US children between the 2003-04 through the 2009-10 seasons.

Results: The average annual rates (per 100,000) of influenza and RSV-associated hospitalizations for any cause excluding asthma (ICD-9 code 493) in the principal diagnosis with a respiratory cause (ICD-9 codes 460-519) in the diagnosis (either principal or secondary) were 173.1(95%CI (134.3,212.6)) (influenza) vs. 2336(2209,2463) (RSV) (age <1y); 77.9(51.1,105.2) vs. 625.7(540,711.5) (age 1y); 56.3(40.9,72.2) vs. 324.2(275.1,375) (age 2y); 44.6(32.9,56.6) vs. 172.2(134.2,211) (age 3y); 36.4(27.7,45.3) vs. 89.7(60.6,117.8) (age 4y); 36.6(29.3,43.9) vs. 62.3(38.8,86.4) (ages 5-6y); 20.5(16.3,24.6) vs. 30.2(16.1,43.9) (ages 7-11y); and 16.1(10.6,21.7) vs. 36.9(17.8,55.9) (ages 12-17y). The rates of RSV-associated hospitalizations with asthma as a secondary (non-principal) diagnosis were 162.6(143.6,181.6) (age <1y), 144.7(120.8,168.2) (age 1y), 99.5(82.3,116.8) (age 2y), 60.9(46.4,75.3) (age 3y), 23.5(10.7,36.4) (age 4y), 16.1(5.1,27) (ages 5-6y), 11.4(4.7,18.3) (ages 7-11y), and 15.8(4.9,26.7) (ages 12-17y).

Conclusions: The estimated rates of RSV-associated hospitalizations in young children are high, with those rates declining rapidly with age. We hope that such estimates could provide one of the ingredients for evaluating the potential impact of RSV vaccine administration in young children. Additionally, the estimated rates of RSV-associated hospitalizations with asthma as a secondary diagnosis in young children, combined with data on the prevalence of asthma in those age groups suggest very high rates of RSV-associated hospitalizations in young children with asthma. Those children may be considered as potential target groups for RSV-related mitigation efforts.

## Introduction

There is uncertainty about the burden of severe outcomes associated with the circulation of respiratory viruses such as RSV and influenza in different subgroups of children defined by age, as well as age and presence of underlying health conditions. Community studies (e.g. [1-3]) reveal various important aspects of the epidemiology of respiratory viruses, such as the high rates of RSV-associated hospitalizations in infants (particularly pre-term infants [1,2]) and young children, and the frequency of underlying health conditions among the severe outcomes (e.g. the high prevalence of asthma among hospitalized children infected with RSV, [1]), etc. At the same time, such studies are subject to limitations such as under-detection of cases of infection, and sample size considerations. Hospital discharge data under-represent the contribution of the respiratory viruses in the diagnoses; moreover, some of those diagnoses may be subject to misspecification of the viral etiology. Attempts at rectifying this (e.g. by attributing certain proportions of hospitalizations with certain diagnoses to respiratory viruses, [4]) are subject to uncertainty. Statistical inference (e.g. [5-8]) is a common approach for estimating the contribution of respiratory viruses to the burden of severe outcomes. However, various assumptions underlying statistical models are uncertain [9,10]; moreover, when one varies the structure of inference models applied to the same data, the resulting goodness-of fit, and the estimates of the contribution of respiratory viruses vary as well ([11,12], Supporting Information). Overall, uncertainty about the burden of severe outcomes associated with respiratory viruses limits the design of mitigation efforts and contributes to the fact that current options for prevention and treatment of infections associated with viruses other than influenza are very limited. Major efforts are currently underway towards the development RSV vaccines, with over 60 candidate vaccines being in different stages of development in 2016 [13-15]. At same time, partly owing to the uncertainty about the severe outcome burden associated with RSV, target groups for those vaccines are uncertain.

In our earlier work [11,12], we have introduced a new method for estimating the burden of severe outcomes associated with influenza and RSV. That method addresses several limitations of previously employed inference methods including utilization of proxies for the incidence of RSV and influenza (sub)types (A/H3N2, A/H1N1, and B) that are expected to be linearly related to the population incidence of those viruses, a relationship that is more biologically plausible [9,10]; utilization of the flexible model for the baseline of severe outcomes not associated with influenza and RSV; application of a bootstrap method for the confidence bounds on the estimates of the severe outcome burden associated with influenza and RSV that accounts for auto-correlation in the time series for the noise; and other constructs. In this paper we apply these methods [11,12] to data on US hospitalizations between the 2003-04 through the 2009-10 seasons. We estimate the rates of influenza and RSV-associated hospitalizations for several diagnoses in various age-groups of children. Our main goals are to provide a granular evaluation (by year of age) of the RSV-associated hospitalization burden in the young children, and to examine the finding in [1] about the high prevalence of asthma among hospitalized children infected with RSV.

**Methods**

*Data*
We used weekly data on US hospitalizations with different diagnoses in different age groups of children (<1y,1y,2y,3y,4y,5-6y,7-11y,12-17y). For our analyses, we included 22 states for which data on hospitalizations in infants with RSV in the principal diagnosis exhibited consistency throughout the study period (see next subsection). Those 22 states constituted approximately 63.2% of the US population during the study period ("AR" "CA" "CT" "GA" "HI" "IL" "IN" "MD" "MN" "NC" "NE" "NJ" "NV" "NY" "OH" "OR" "TN" "TX" "VA" "VT" "WA" "WI"). Incidence proxies for the major influenza (sub)types are derived using the US CDC influenza surveillance data [16].

*Inference Scheme*
Our methodology is based on the framework developed in [12].

Each season was defined as a period between calendar week 27 of a given year, and calendar week 26 of the following year.

For the influenza incidence proxies, we utilize the US CDC surveillance data [16]. We combine the weekly state-specific data on the percent of medical consultations in the US CDC Outpatient Illness Surveillance Network (ILINet) that were for influenza-like illness with data on testing of respiratory specimens for the major influenza (sub)types (A/H3N2, A/H1N1, and influenza B) to define state-specific weekly proxies for the incidence of each of the major influenza (sub)type:

*Weekly influenza (sub)type incidence proxy =* (1)

*= % Consultations for ILI \* % Specimens testing positive for the (sub)type*

Our analyses pertain to the collection of 22 US states described in the Data subsection, and the incidence proxy for this collection of states is defined as the sum of state-specific incidence proxies weighted by the state populations (with state populations during various weeks estimated through linear interpolation (in time) for the yearly, July 1$^{st}$ estimates in [28]) .

We consider the weekly incidence proxy for RSV to be the rate of hospitalizations with the principal diagnosis of RSV among children aged <1y during that week. We use hospitalization rates in infants because we seek a proxy proportional to RSV incidence in the population, and there is an apparent upward trend in annual rates of hospitalizations with the principal diagnosis of RSV in all age groups above 2 years of age, likely due to changes in testing/diagnostic practices, with no apparent trend in the rates of hospitalization with a principal diagnosis of RSV in infants. We note that in our earlier work [12], we used rates of RSV bronchiolitis in infants as a proxy for RSV incidence because of evidence in the literature (e.g. [17]) that the vast majority of bronchiolitis diagnoses were supported by laboratory tests. For this study, we were not in possession of data on RSV bronchiolitis in infants. At the same time, practices for testing for viral infections following a bronchiolitis diagnosis have changed over time, and laboratory testing is no longer recommended in the US [18]. We also note that rates of hospitalizations in infants with RSV in the principal diagnosis may not be proportional to RSV incidence rates because presence of RSV in the principal diagnosis need not be supported by a laboratory test. Correspondingly, diagnostic practices may vary by place (e.g. hospitalizations in infants with symptoms that are often associated with RSV infections, such as wheezing, may result in RSV diagnoses without confirmation by laboratory testing more often in some places compared to others). Additionally, temporal trends in diagnostic practices (e.g. ones resulting from changes in testing practices) may also be present in certain locations. We've tried to adjust for such potential biases by excluding states with an apparent trend in the rates of hospitalizations in infants with RSV in the principal diagnosis, as well as states for which annual rates of hospitalizations in infants with RSV in the principal diagnosis are noticeably higher than in most states.

We note, as we did in [12], that the relation between an influenza incidence proxy and the rate of associated hospitalizations in a given age group may change in time due to a variety of factors. In particular, changes in the age distribution for different influenza-associated outcomes, e.g. the rise in the proportion of school-age children during seasons when novel influenza strains were circulating are known to have taken place. For example, the 2003-04 influenza season was driven by a novel (Fujian) A/H3N2 strain. Results in [8] suggests that the rates of excess ED visits for school-age children compared to pre-school children were higher during the 2003-04 season compared to the 2004-05 season. Similarly, the emergence of the pandemic A/H1N1 influenza strain in 2009 resulted in exceptionally high rates of pediatric influenza infections [19], and the ratios between the rates of A/H1N1-associated hospitalizations in different subgroups of children and the A/H1N1 incidence proxy are potentially different for the pandemic strain compared to the seasonal A/H1N1 strain. Based on those considerations, we split the A/H3N2 incidence proxy into two, namely H31 equaling the A/H3N2 proxy for the 2003-04 season, and being zero afterwards, and H32, being zero for the 2003-04 season, and equaling the A/H3N2 incidence proxy afterwards. Similarly, we split the A/H1N1 incidence proxy into two, reflecting A/H1N1 incidence before and after the 2009 pandemic.

The outcomes we consider are hospitalizations with different diagnoses in different age groups of children. The four types of hospitalization diagnoses used in our analyses are: diseases of the respiratory system excluding asthma (ICD 9 codes 460-519 excluding 493); pneumonia and influenza (ICD9 codes 480-488); any code excluding asthma in the principal diagnosis with a respiratory code in the principal or secondary diagnoses; and asthma as a secondary (non-principal) diagnosis (see the next sub-section). For each outcome of interest, weekly rates for this outcome during the study period are regressed linearly against the incidence proxies for influenza, RSV, as well as temporal trend (modeled by a low degree polynomial in time) and a seasonal baseline (modeled by cubic splines in the calendar week with knots at every fourth week and with annual periodicity) [11,12]. If $O(t)$ is the weekly rate for the outcome of interest on week $t$, and $V_i(t)$ are the incidence proxies for the various influenza (sub)types (split into time periods as described above) and RSV, the model structure is

$$O(t) = \sum_i \beta_i \cdot V_i(t) + Baseline + Trend + Noise \quad (2)$$

To account for the autocorrelation structure in the noise, a bootstrap method devised in [11] is used to prescribe the confidence bounds for the model's estimates. Estimates in [12] suggest that several alternative inference methods (such as maximal likelihood estimation under the AR(1) assumption on the noise) yield very similar results.

*Asthma-related hospitalizations*
While we've estimated the rates of influenza and RSV-associated hospitalizations with asthma as a secondary (non-principal) diagnosis, hospitalizations for asthma (as the principal diagnosis) were excluded from the analysis. Briefly, this was done because asthma hospitalizations do not accord well with the structure of the inference model (eq. 2), as suggested both by visual inspection and inconsistencies in the estimates provided by the models.

If one assumes that for RSV-associated hospitalizations in children with asthma in the secondary diagnosis, asthma was present in those children before the RSV infection (see the penultimate paragraph of the Discussion) then the rates of RSV-associated hospitalizations with asthma in the secondary diagnosis in children with asthma in a given age group equal the corresponding rates in all children in that age group divided by the prevalence of asthma in that age group.

## Results

Tables 1 and 2 present the estimates of the average annual rates of hospitalizations associated with influenza and RSV for various diagnoses in select age subgroups of children. Table 1 exhibits the results for hospitalizations for respiratory causes (excluding asthma) in the principal diagnosis, as well as for any cause excluding asthma in the principal diagnosis with a respiratory cause in either the principal or secondary diagnosis. Table 2 presents the corresponding results for hospitalizations for Pneumonia and Influenza (P&I) in the principal diagnosis, as well for hospitalizations with asthma as a secondary (non-principal) diagnosis.

Given the low prevalence of asthma in young children (4.7% among children under the age of 5y, [20]), our estimates suggest that the rates of RSV-associated hospitalizations in young children with asthma are very high. In fact this already follows from the estimates of the rates of RSV-associated hospitalizations with asthma as a secondary diagnosis in young children (Methods), with further contribution represented by RSV-associated hospitalizations with asthma as the principal diagnosis, as well as RSV-associated hospitalizations in children with asthma for which asthma does not appear in the diagnosis – see Discussion.

For any studied age group and diagnosis (save for P&I hospitalizations in children aged 7-11y), the estimated rates of RSV-associated hospitalizations are higher than the estimated rates of influenza-associated hospitalizations. The estimated rates of RSV-associated hospitalizations for any principal diagnosis (excluding asthma) with a respiratory cause in the diagnosis (either principal or secondary, Table 1) are high in young children, though they decline rapidly with age. For the more restricted category of respiratory causes in the principal diagnosis, the rates of RSV-associated hospitalizations were somewhat lower (Table 1); the corresponding rates of RSV-associated hospitalizations with P&I in the principal diagnosis were further lower, particularly for infants aged <1y (Table 2).

| Age group / Diagnosis | Respiratory cause (excluding asthma) in the principal diagnosis | | Hospitalizations (excluding asthma) with respiratory cause in principal or secondary diagnosis | |
|---|---|---|---|---|
| | Flu | RSV | Flu | RSV |
| <1y | 134.6 (107,162.9) | 2106 (2017,2193) | 173.1 (134.3,212.6) | 2336 (2209,2463) |
| 1y | 62.7 (38.6,87.6) | 549.2 (475.2,625.2) | 77.9 (51.1,105.2) | 625.7 (540,711.5) |
| 2y | 41.8 (29,54.9) | 264.6 (222.9,305.4) | 56.3 (40.9,72.2) | 324.2 (275.1,375) |
| 3y | 32.3 (23.1,41.7) | 142.7 (113.9,172.5) | 44.6 (32.9,56.6) | 172.2 (134.2,211) |
| 4y | 25.6 (18.8,32.4) | 63.9 (42,85.4) | 36.4 (27.7,45.3) | 89.7 (60.6,117.8) |
| 5-6y | 27.2 (21.6,32.7) | 42.2 (24.2,59.6) | 36.6 (29.3,43.9) | 62.3 (38.8,86.4) |
| 7-11y | 14.6 (12.2,17) | 18 (10.4,25.4) | 20.5 (16.3,24.6) | 30.2 (16.1,43.9) |
| 12-17y | 9.8 (8.4,11.3) | 14.6 (9.7,19.4) | 16.1 (10.6,21.7) | 36.9 (17.8,55.9) |

**Table 1:** Average annual rates of hospitalization per 100,000 associated with influenza and RSV for respiratory causes excluding asthma in the principal diagnosis (ICD-9 codes 460-519 excluding 493), and for any principal diagnosis excluding asthma with a respiratory cause in the principal or secondary diagnoses in select groups of US children, 2003-04 through 2009-10 seasons.

| Age group / Diagnosis | Pneumonia and Influenza in the principal diagnosis | | Asthma as a secondary (non-principal) diagnosis | |
|---|---|---|---|---|
| | Flu | RSV | Flu | RSV |
| <1y | 105.9 (84.4,127.9) | 340.7 (276.5,403.5) | 5.1 (-0.7,11) | 162.6 (143.6,181.6) |
| 1y | 55.6 (43.2,68.4) | 241.2 (200.8,281.8) | 10 (2.8,17.3) | 144.7 (120.8,168.2) |
| 2y | 35.1 (27.1,43) | 151.7 (125.8,178.1) | 9.2 (3.9,14.5) | 99.5 (82.3,116.8) |
| 3y | 28.1 (20.4,36.1) | 94.8 (70.6,119.4) | 8 (3.8,12.3) | 60.9 (46.4,75.3) |
| 4y | 21.8 (16.6,27.2) | 44.9 (27.9,61.7) | 6.3 (2.6,10.2) | 23.5 (10.7,36.4) |
| 5-6y | 21.2 (16.5,26) | 29.6 (14,44.5) | 7.2 (3.9,10.5) | 16.1 (5.1,27) |
| 7-11y | 11.2 (9.3,13.1) | 10.9 (4.9,16.9) | 5 (3.1,7) | 11.4 (4.7,18.3) |
| 12-17y | 7.9 (7,8.8) | 8 (5,11) | 2.2 (-1,5.4) | 15.8 (4.9,26.7) |

**Table 2:** Average annual rates of hospitalization per 100,000 associated with influenza and RSV for P&I (ICD9 codes 480-488), and asthma as a secondary (non-principal) diagnosis in select groups of US children, 2003-04 through 2009-10 seasons.

**Discussion**

While major efforts are currently under way towards the development of RSV vaccines [13-15], the design of the corresponding vaccination strategies (and other mitigation efforts) is limited by the uncertainty about the severe outcome burden associated with RSV. In this paper we apply the previously developed methodology [11,12] to estimate the rates of influenza and RSV-associated hospitalizations with various diagnoses in different age subgroups of US children between 2003-2010. Our estimates suggest that the rates of RSV-associated hospitalizations in young children are high, though they decline rapidly by year of age. Additionally, our results suggest that the rates of RSV-associated hospitalizations in young children with asthma are extremely high, making those children potential targets for related mitigation efforts. Indeed, the US CDC estimates suggest prevalence of 4.7% for asthma in children under the age of 5y in 2015 [20]. Combining this with the estimates in Table 2 (Methods) tells that the rates of RSV-associated hospitalizations with asthma in the secondary diagnosis per 100,000 children with asthma are 3460(3055,3864) (age <1y); 3079(2570,3579) (age 1y); 2117(1751,2485) age (2y); 1296(987,1602) (age 3y), etc.; those estimates are in addition to RSV-associated hospitalizations with asthma in the principal diagnosis, or not appearing in the diagnosis. We note that estimates of the prevalence of asthma in young children may be subject to under-detection. Even if the true prevalence of asthma in young children is higher than the estimates in [20], closer to the 10% rate for older children [20], this still suggests that for children under the age of 2y with asthma, the annual rates of RSV-associated hospitalizations with asthma in the secondary diagnosis alone are around 1,500/100,000, with the corresponding rate being around 1,000/100,000 for children aged 2y, etc.

While the connection between RSV infection in early childhood and the subsequent risk of developing asthma has received a lot of attention in the literature [21,25-27], the contribution of RSV to the hospitalizations burden for children with asthma is less studied [22]. The results in [1] suggest high frequency of RSV infections in hospitalized children under the age of 5y, with a high proportion of RSV-infected hospitalized children having asthma. Other studies have found high frequency of RSV infections in hospitalized young children with underlying health conditions, including asthma [23,24]. Our estimates of the rates of RSV-associated hospitalizations with asthma as a secondary diagnosis support the results of those studies, providing further evidence about the very high rates of RSV-associated hospitalizations in young children with asthma.

Our paper has some limitations. One of them is the accuracy of the proposed regression framework [11,12]. Spikes in hospitalization rates during certain influenza seasons correspond visually to major influenza epidemics (as suggested by the incidence proxies that we utilize), providing additional support for the validity of our inference method for influenza-associated hospitalization rates. Inclusion of RSV into the model can be more problematic due to the high periodicity of its circulation and the difficulty of separating its contribution from the annual baseline of hospitalization rates not associated with influenza and RSV. Additionally, we've used rates of hospitalizations in infants with RSV in the principal diagnosis as proxy for RSV incidence, while presence of RSV in the principal diagnosis need not be supported by a laboratory test. We've tried to remedy this by excluding states where rates of hospitalization in infants with RSV in the principal diagnosis were noticeably higher than average (which might suggest admixture from other etiologies in that category of hospitalizations) from the analyses (see Methods). While the resulting confidence bounds for the contribution of RSV to many of the hospitalization categories that we've considered are reasonably tight (Tables 1 and 2), further work is needed to evaluate the accuracy of our inference methodology. It is not entirely clear whether for RSV-associated hospitalizations in children with asthma in the secondary diagnosis, asthma was present before the RSV infection, or developed between the RSV infection and hospitalization. We assume that for the overwhelming majority of those children, the former is the case. While the connection between severe RSV infection and the subsequent development of asthma was studied by various authors [25-27], this development doesn't appear to take place immediately. It has been suggested ([26]) that a severe RSV episode initiates a series of immunologic events that eventually result in asthma. Finally, while we estimated the rates of influenza and RSV-associated hospitalizations with asthma as a secondary (non-principal) diagnosis, we could not estimate the corresponding rates for hospitalizations with asthma as a principal diagnosis (see Methods). Nonetheless, our results for hospitalizations with asthma as a secondary diagnosis already suggest very high rates of RSV-associated hospitalizations for young children with asthma.

We believe that despite those limitations, our work provides a granular assessment of the hospitalization burden associated with influenza and RSV in young children that may represent one of the ingredients for evaluating the impact of RSV vaccination in young children. Additionally, our results suggest very high rates of RSV-associated hospitalization in young children with asthma. Those children may be considered as potential target groups for RSV-related mitigation efforts, such as palivizumab prophylaxis.